\documentclass[preprint,showpacs,preprintnumbers,amsmath,amssymb]{revtex4}
\usepackage{graphicx}% Include figure files

\begin{document}

\title{Electrostatic models of electron-driven proton transfer \\ across a lipid membrane}

\author{ Anatoly  Yu. Smirnov$^{1,2}$, Lev G. Mourokh$^{3}$, and Franco Nori$^{1,2}$}

\affiliation{ $^1$ Advanced Science  Institute, RIKEN, Wako-shi, Saitama, 351-0198, Japan \\
$^2$ Physics Department, The University of Michigan, Ann Arbor, MI 48109-1040, USA,\\
$^3$ Department of Physics, Queens College, The City University of New York, Flushing, New York 11367, USA}

\date{\today}

\begin{abstract}
{We present two models for electron-driven uphill proton transport across lipid membranes, with the electron energy converted to the proton
gradient via the electrostatic interaction. In the first model, associated with the \emph{cytochrome c oxidase} complex in the inner
mitochondria membranes, the electrostatic coupling to the site occupied by an electron lowers the energy level of the proton-binding site,
making the proton transfer possible. In the second model, roughly describing the redox loop in a nitrate respiration of \emph{E. coli} bacteria,
an electron displaces a proton from the negative side of the membrane to a shuttle, which subsequently diffuses across the membrane and unloads
the proton to its positive side. We show that both models can be described by the same approach, which can be significantly simplified if the
system is separated into several clusters, with strong Coulomb interaction inside each cluster and weak transfer couplings between them. We
derive and solve the equations of motion for the electron and proton creation/annihilation operators, taking into account the appropriate
Coulomb terms, tunnel couplings, and the interaction with the environment. For the second model, these equations of motion are solved jointly
with a Langevin-type equation for the shuttle position. We obtain expressions for the electron and proton currents and determine their
dependence on the electron and proton voltage build-ups, on-site charging energies, reorganization energies, temperature, and other system
parameters. We show that the quantum yield in our models can be up to 100\% and the power-conversion efficiency can reach 35\%.}
\end{abstract}

\pacs{82.39.Jn, 87.16.A-, 73.63.-b}

\maketitle

\section{Introduction}

Every living organism obtains the energy needed for its survival from the outside world. This energy can be in the form of sunlight or food; but
in both cases it is unstable and cannot be utilized directly, so several energy-conversion steps are necessary. One of the most widely used
intermediate forms for energy storage is the electrochemical proton gradient across lipid membranes, such as the inner mitochondrial membranes
or plasma membranes in bacteria. To achieve and maintain this proton gradient, nature employs several different types of electron- or
light-driven systems, where the energy of high-energetic electrons or absorbed photons is used for the energetically-uphill proton transfer from
the negative ($N$) to the positive ($P$) sides of the membrane.

Here we discuss two mechanisms of energy conversion from the highly unstable electronic form of energy to the proton gradient, namely, proton
pumps and redox loops \cite{Alberts02,Nicholls02}. Both mechanisms rely on the electrostatic interaction between electrons and protons, although
the specific details of the proton pumps and the redox loops look very different. For example, in a proton pump, such as \emph{cytochrome c
oxidase}, electrons move mainly along the membrane, whereas protons move across the membrane, which results in an accumulation of the positive
charge on the $P$-side and in the generation of a proton-motive force (PMF) \cite{WV07,BelPNAS07,Kim07,JCPCCO}. In the redox-loop mechanism of
PMF generation, taking place in the nitrate respiratory chain of \emph{E. coli} bacterium, the \emph{neutral shuttle}, carrying \emph{both}
protons and electrons, crosses the membrane. Here, the charge accumulation occurs when electrons cross the membrane, just before embarking on
the shuttle, and right after unloading from the shuttle \cite{Mitch76,Jormakka02,Bertero03,Gennis05,PRERedoxLoop,JCPPhoto}. It should be noted
that the proton pump operating in the \emph{cytochrome c oxidase} has no essential mechanically-moving parts, whereas the redox-loop mechanism
is impossible without the molecular shuttle diffusing between the negative and the positive sides of the lipid membrane.

In general, the treatment of the electron and proton transfer events is extremely difficult because the total number of the occupation states
increases exponentially with the number of the electron- and proton-binding sites, when all of them are electrostatically coupled. In the
present work, however, we show that both above-mentioned mechanisms of the transmembrane proton translocation can be described with a similar
mathematical model, taking into account the Coulomb interaction between \emph{one} electron- and \emph{one} proton-binding sites only, and
neglecting electrostatic couplings to other sites. It is necessary to have at least three redox sites and three proton-binding sites in order to
obtain a proton pumping effect and suppress a reverse flow of protons from the $P$-side to the $N$-side of the membrane. In the absence of
strong Coulomb interaction between all sites, there is no need to introduce a complete set of electron and proton occupation states (as was done
in our previous works, Refs.~\cite{JCPCCO,PRERedoxLoop,JCPPhoto,PREForster}), which grows exponentially with the number of sites. Instead, we
now divide the whole system into clusters of strongly coupled sites. These clusters are described by their own set of occupation states, and the
total number of the states in the system is equal to the sum (not the product!) of the states in the clusters. The clusters are weakly coupled
by electron tunneling terms and by proton transfer amplitudes, so that transitions between the clusters can be considered within perturbation
theory. While in this work we present quite simple models, similar approaches can be applied to much more complicated biological systems, such
as Photosystem II and the whole respiratory chain in the inner mitochondrial membrane \cite{Nicholls02}.

The quantum yield for the two models analyzed in this paper can be about 1. Why such a high quantum yield?  This can be explained from the fact
that, in order to be transferred through the system, an electron needs to loose its energy. This cannot be done via the environment because the
reorganization energy is not large enough. Consequently, electron transport occurs with the assistance of protons gaining this energy and being
transferred to the positive side of the membrane. Thus, the transfer of a single electron is accompanied by the transfer of a single proton and
the corresponding currents are equal, which results in an almost perfect quantum yield.

\section{Model}
We consider a physical model describing an electron-coupled translocation of protons from the negative ($N$) to the positive ($P$) side of a
membrane. The model consists of an interaction site, $Q = \{Q_e,\,Q_p\}$, containing a single electron level with energy $\varepsilon_Q$ and a
single proton energy level characterized by the energy $E_Q$. We also introduce two electron sites, $L$ and $R$, coupled to the electron site
$Q_e$, and two proton sites, $A$ and $B$, coupled to the proton site $Q_p$ (Fig.~1). The electron site $L$ is coupled to the electron source
$S$, and the site $R$ is connected to the electron drain $D$. The proton site $A$ is coupled to the proton reservoir $N$ (the negative side of
the membrane), and the site $B$ is coupled to the positive side of the membrane (proton reservoir $P$).

\subsection{Hamiltonian}

The Coulomb interaction between an electron and a proton, both located on the central site $Q$, is described by the energy $u_0$, so that the
Hamiltonian of the site $Q$ has the form
\begin{equation}
H_Q = \varepsilon_Q n_Q + E_Q N_Q - u_0 n_Q N_Q, \label{HQ}
\end{equation}
where $n_Q = a_Q^\dag a_Q$ is the electron population of the site Q, and $N_Q = b_Q^\dag b_Q$ is the proton population of this site. Electrons
are described by the Fermi-operators $a_{\sigma}$, and protons are characterized by the Fermi-operators $b_{\alpha}$ with $\sigma = L,Q_e,R$ and
$\alpha = A,Q_p,C,$ and with the corresponding populations $n_{\sigma} = a_{\sigma}^\dag a_{\sigma},\; N_{\alpha} = b_{\alpha}^\dag b_{\alpha}.$

The contribution of the electron sites $L,R$ and the proton sites $A,B$ to the total Hamiltonian of the system is described by the term
\begin{equation}
H_0 = \varepsilon_L n_L + \varepsilon_R n_R  + E_A N_A + E_B N_B, \label{H0}
\end{equation}
where $\varepsilon_L, \varepsilon_R$ are the energy levels of the electron sites $L$ and $R$, and $E_A, E_B$ are the energies of the
proton-binding sites $A$ and $B$.

The strongly-interacting electron and proton sites $Q_e$ and $Q_p$ form a single (interaction) cluster, whereas the sites $L,R$ and $A,B$
separately form other four (peripheral) clusters. The cluster $Q$ can be characterized by the vacuum (empty) state and by three additional
occupation states, or, equivalently, by the average electron and proton populations, $\langle n_Q\rangle $ and $\langle N_Q\rangle $,
complemented by the correlation function, $K = \langle n_Q N_Q\rangle.$ The other electron and proton clusters are described by the
corresponding average occupations, $\langle n_L\rangle ,\;\langle n_R\rangle $ and $\langle N_A\rangle ,\; \langle N_B\rangle $. For six
electron and proton-binding sites we should have $2^6 = 64$ occupation states. However, with the cluster approach, the system can be completely
described by only seven functions: $\langle n_Q\rangle, \langle N_Q \rangle,  K $ (for the interaction cluster), and $\langle n_L\rangle ,
\langle n_R\rangle, \langle N_A\rangle , \langle N_B\rangle $ (for the peripheral clusters). Previously, we applied a similar approach to
analyze quantum transport problems in nanomechanical systems \cite{RobertPRB}.

\subsubsection{Electron and proton transitions} The electron tunneling  Hamiltonian between the site $Q$ and the sites $L$ and $R$ is given by
\begin{equation}
H_e = - \Delta_L\, a_L^\dag\, a_Q - \Delta_R\, a_R^\dag \,a_Q + {\rm H.c.}, \label{He}
\end{equation}
whereas the $A$-$Q$ and $B$-$Q$ proton transitions are described by the term
\begin{equation}
H_p = - \Delta_A \,b_A^\dag\, b_Q - \Delta_B \,b_B^\dag \,b_Q + {\rm H.c.}. \label{Hp}
\end{equation}
Here $\Delta_L, \Delta_R$ are the electron tunneling coefficients, and $\Delta_A, \Delta_B$ are the proton transfer amplitudes. In the case of a
movable interaction site, e.g., when  the electron and proton sites $Q$ are located on the shuttle (quinone/quinol), the amplitudes $\Delta_L,
\Delta_R$ and $\Delta_A, \Delta_B$ depend on the position $x$ of the shuttle.

The $S$-lead serves as a source of electrons, and the $D$-lead works as an electron drain. The coupling to these leads is characterized by the
Hamiltonian
\begin{equation}
H_{LR} = -\sum_k t_{kS}\, c_{kS}^\dag\, a_L - \sum_k t_{kD} \,c_{kD}^\dag \,a_R + {\rm H.c.}  \label{HLR}
\end{equation}
The proton transitions between the $N$-side of the membrane and the site $A$ and between the $P$-side of the membrane and the site $B$ are
described by the Hamiltonian
\begin{equation}
H_{AB} = - \sum_q T_{qN} \,d_{qN}^\dag\, b_A - \sum_q T_{qP}\,d_{qP}^\dag\, b_C + {\rm H.c.} \label{HAB}
\end{equation}
Here $c_{kS}, c_{kD}$ are Fermi operators of the electron reservoirs $S$ and $D$, and $d_{qN}, d_{qP}$ are the Fermi operators of protons in the
reservoirs $N$ and $P$. The electron reservoirs $S$ and $D$ have the Hamiltonian
\begin{equation}
H_{SD} = \sum_k ( \varepsilon_{kS} \,c_{kS}^\dag\, c_{kS} + \varepsilon_{kD}\, c_{kD}^\dag \,c_{kD}), \label{HSD}
\end{equation}
and are characterized by the Fermi distributions $f_S(\varepsilon_{kS}), f_D(\varepsilon_{kD})$ with the corresponding electrochemical
potentials $\mu_S$ and $\mu_D$. For the proton reservoirs $N$ and $P$ we have the Hamiltonian
\begin{equation}
H_{NP} = \sum_q ( E_{qN}\, d_{qN}^\dag\, d_{qN} + E_{qP} \,d_{qP}^\dag\, d_{qP} ), \label{HNP}
\end{equation}
with the Fermi distributions $F_N(E_{qN})$ and $F_P(E_{qP})$ and the proton electrochemical potentials $\mu_N$ and $\mu_P$.

\subsubsection{Environment}
The interaction of the electron-proton system with the protein environment, which is described as a sum of independent oscillators
\cite{Krish01}, is characterized by the Hamiltonian
\begin{eqnarray}
H_{\rm env} = \sum_j \frac{p_j^2}{2 m_j} + \sum_j \frac{m_j\omega_j^2}{2} \left(x_j - \sum_{\sigma} x_{j\sigma} n_{\sigma} - x_{jS} \sum_k
c_{kS}^\dag c_{kS} - x_{jD} \sum_k
c_{kD}^\dag c_{kD} - \right. \nonumber\\
\left. \sum_{\alpha} X_{j\alpha} N_{\alpha} - X_{jN}\sum_q d_{qN}^\dag d_{qN} - X_{jP}\sum_q d_{qP}^\dag d_{qP} \right)^2, \label{Henv}
\end{eqnarray}
where $n_{\sigma} = a_{\sigma}^\dag a_{\sigma}$ is the population of the electron site $\sigma$ ($\sigma = L,Q,R$), $ N_{\alpha} =
b_{\alpha}^\dag b_{\alpha}$ is the population of the proton site $\alpha$ ($\alpha = A,Q,B$). The constants $x_{j\sigma}, x_{jS}, x_{jD}$
determine the electron coupling to the environment, and the parameters $X_{j\alpha}, X_{jN}, X_{jP}$ describe the proton-environment
interaction.

With the unitary transformation,
\begin{eqnarray}
{\cal U} = \exp \bigg[ - i \sum_j p_j \bigg(\sum_{\sigma} x_{j\sigma} n_{\sigma} + x_{jS} \sum_k c_{kS}^\dag c_{kS} + x_{jD} \sum_k
c_{kD}^\dag c_{kD} +  \nonumber\\
\sum_{\alpha} X_{j\alpha} N_{\alpha} + X_{jN}\sum_q d_{qN}^\dag d_{qN} + X_{jP}\sum_q d_{qP}^\dag d_{qP} \bigg) \bigg], \label{UT}
\end{eqnarray}
the environment Hamiltonian can be rewritten as
\begin{equation}
H_{\rm env} = \sum_j\left( \frac{p_j^2}{2 m_j} + \frac{m_j \omega_j^2 x_j^2}{2}\right), \label{Henv2}
\end{equation}
whereas the Hamiltonians $H_e$ and $H_p$ acquire the stochastic phase factors:
\begin{equation}
H_e = - \Delta_L\, e^{i\xi_{L}}\, a_L^\dag\, a_Q - \Delta_R\, e^{i \xi_{R}} \,a_R^\dag\, a_Q + {\rm H.c.}, \label{He2}
\end{equation}
and
\begin{equation}
H_p = - \Delta_A \,e^{i\xi_{A}}\, b_A^\dag\, b_Q - \Delta_B\, e^{i\xi_{B}}\,b_B^\dag \,b_Q + {\rm H.c.} \label{Hp2}
\end{equation}
with the phases $$ \xi_{L} = \sum_j p_j (x_{jL} - x_{jQ}),$$ $$ \xi_{R} = \sum_j p_j (x_{jR} - x_{jQ}),$$ and $$\xi_{A} = \sum_j p_j (X_{jA} -
X_{jQ}),$$ $$ \xi_{B} = \sum_j p_j (X_{jB} - X_{jQ}).$$ For simplicity, we assume that there are no phase shifts for the electron transitions
between the electron source $S$ and the site $L$, and the electron drain $D$ and the site $R$, so that $x_{jS} = x_{jL},$ and $x_{jD} = x_{jR}$,
with the same assumption for the $N$-$A$ and $P$-$B$ proton transitions, $X_{jN} = X_{jA}$ and $X_{jP} = X_{jB}.$

\subsection{Rate equations}
The time evolution of the electron operators $n_{\sigma}$ is determined by the Heisenberg equations:
\begin{eqnarray}
\dot{n}_L = i \Delta_L \,e^{i\xi_{L}}\, a_L^\dag \,a_Q - i \sum_k t_{kS}\,
c_{kS}^\dag\, a_L + {\rm H.c.}, \nonumber\\
\dot{n}_R = i \Delta_R \,e^{i\xi_{R}} \,a_R^\dag \,a_Q - i \sum_k t_{kD} \,c_{kD}^\dag \,a_R + {\rm H.c.}, \label{nLRH}
\end{eqnarray}
and
\begin{equation}
\dot{n}_Q = -i \Delta_L \,e^{i\xi_{L}} \,a_L^\dag\, a_Q - i \Delta_R \,e^{i\xi_{R}}\, a_R^\dag \,a_Q + {\rm H.c.} \label{nQH}
\end{equation}
For the proton populations $N_{\alpha}$, we derive the similar set of Heisenberg equations,
\begin{eqnarray}
\dot{N}_A = i \Delta_A \,e^{i\xi_{A}}\, b_A^\dag \,b_Q - i \sum_q T_{qN}\,
d_{qN}^\dag \,b_A + {\rm H.c.}, \nonumber\\
\dot{N}_B = i \Delta_B \,e^{i\xi_{B}} \,b_B^\dag\, b_Q - i \sum_q T_{qP} \,d_{qP}^\dag\, b_B + {\rm H.c.} \label{NABH}
\end{eqnarray}
This set should be complemented by the equation for the proton population of the interaction site,
\begin{equation}
\dot{N}_Q = -i \Delta_A \,e^{i\xi_{A}}\, b_A^\dag \,b_Q -i \Delta_B\, e^{i\xi_{B}}\, b_B^\dag \,b_Q + {\rm H.c.}, \label{NQH}
\end{equation}
as well as by the equations for the operators of electron and proton reservoirs,
\begin{eqnarray}
i\, \dot{c}_{kS} &=& \varepsilon_{kS}\, c_{kS} - t_{kS}\, a_L, \nonumber\\
i\, \dot{c}_{kD} &=& \varepsilon_{kD} \,c_{kD} - t_{kD}\, a_R, \label{cSD}
\end{eqnarray}
\begin{eqnarray}
i \,\dot{d}_{qN} &=& E_{qN}\, d_{qN} - T_{qN}\, b_A, \nonumber\\
i\, \dot{d}_{qP} &=& E_{qP} \,d_{qP} - T_{qP}\, b_B. \label{dNP}
\end{eqnarray}

\subsubsection{Contribution of reservoirs to the rate equations}
It follows from Eq.~(\ref{cSD}) that the electron operator $c_{kS}$ can be represented as
\begin{equation}
c_{kS} = c_{kS}^{(0)} - t_{kS} \int dt_1 \langle -i [ c_{kS}^{(0)}(t), c_{kS}^{(0)\dag}(t_1)]_{+}\rangle \, a_L (t_1) \,\theta(t-t_1),
\label{cS}
\end{equation}
where $c_{kS}^{(0)}(t)$ is the free variable of the $S$-lead, and $\theta(t-t_1)$ is the Heaviside step function. Similar expressions take place
for the electron operator $c_{kD}(t)$ and for operators $d_{qN}, d_{qP}$ of the proton reservoirs. For the weak coupling between the reservoir
$S$ and the electron site $L$ we obtain
\begin{equation}
\langle a_L^\dag(t) c_{kS}^{(0)}(t) \rangle = - i t_{kS} \int dt_1 \langle c_{kS}^{(0)\dag}(t_1) c_{kS}^{(0)}(t)\rangle\, \langle [a_L(t_1),
a_L^\dag(t)]_{+}\rangle \,\theta(t-t_1). \label{aLcS}
\end{equation}
Thus,  contribution of the $S$-lead to the evolution of the average electron population $\langle n_L\rangle$ (see Eq.~(\ref{nLRH})) is
determined by the expression
\begin{eqnarray}
i\,\sum_k t_{kS}^* \langle a_L^\dag (t) c_{kS}(t)\rangle = - \sum_k |t_{kS}|^2 \int dt_1 \{ \langle
c_{kS}^{(0)}(t)c_{kS}^{(0)\dag}(t_1)\rangle\, \langle a_L^\dag (t) a_L(t_1)\rangle -\nonumber\\ \langle
c_{kS}^{(0)\dag}(t_1)c_{kS}^{(0)}(t)\rangle\, \langle a_L(t_1) a_L^\dag(t)\rangle\}. \label{aLcS2}
\end{eqnarray}
The correlator $\langle c_{kS}^{(0)\dag}(t_1)c_{kS}^{(0)}(t)\rangle$ is proportional to the Fermi distribution function, $f_S(\varepsilon_{kS})$
of electrons in the reservoir $S$,
\begin{equation}
\langle c_{kS}^{(0)\dag}(t_1)c_{kS}^{(0)}(t)\rangle = f_S(\varepsilon_{kS}) \, e^{-i \varepsilon_{kS}(t-t_1)}, \label{cScS}
\end{equation}
where the Fermi function, $$f_S(\varepsilon) = \left[ \exp\left(\frac{\varepsilon - \mu_S}{T}\right) + 1 \right]^{-1},$$ is characterized by the
electrochemical potential $\mu_S$ and temperature $T$. We assume that the site $L$ is weakly-coupled to the reservoir $S$ and to the site $Q$,
thus, we can use free-evolving operators, $$a_L(t) = e^{-i\varepsilon_L (t-t_1)}\, a_L(t_1),$$ to calculate the corresponding correlation
functions in Eq.~(\ref{aLcS2}), e.g., $$\langle a_L^\dag (t)\, a_L(t_1) \rangle = \langle n_L(t) \rangle \,e^{i\varepsilon_L(t-t_1)}.$$
Introducing the energy-independent rate constant,
\begin{equation}
\gamma_S = 2\pi \sum_k |t_{kS}|^2 \, \delta(\varepsilon_L - \varepsilon_{kS}),
\end{equation} \label{gammaS}
we calculate the contribution of the $S$-lead to the time evolution of the population $\langle n_L\rangle$,
\begin{equation}
i\,\sum_k t_{kS}^* \, \langle a_L^\dag (t) c_{kS}(t)\rangle + {\rm H.c.} = \gamma_S \, [ f_S(\varepsilon_L) - \langle n_L\rangle ].
\label{aLcS3}
\end{equation}
The same analysis can be applied for a calculation of contributions of the electron lead $D$ and the proton leads $N$ and $P$ to the
corresponding populations $\langle n_R\rangle$ and $\langle N_A\rangle, \langle N_P\rangle.$ The proton transfer rates between the sites $A$ and
$C$ and the negative and positive sides of the membrane, respectively, are determined by the coefficients $\Gamma_N$ and $\Gamma_P$ where, e.g.,
\begin{equation}
\Gamma_N = 2\pi \sum_q |T_{qN}|^2 \, \delta(E_A - E_{qN}). \label{GammaN}
\end{equation}

\subsubsection{Contribution of site-to-site tunneling to the rate equations}
To calculate a contribution of the $L$-$Q$ tunneling to the evolution of the populations $\langle n_L\rangle$ and $\langle n_Q\rangle$, we start
with the amplitude $a_Q$, which obeys the equation
\begin{equation}
i \dot{a}_Q = \varepsilon_Q \, a_Q - u_0 \, N_Q \,a_Q - \Delta_L^* \,e^{-i\xi_{L}}\, a_L - \Delta_R^* \,e^{-i\xi_{R}}\, a_R. \label{aQ1}
\end{equation}
In the case of weak $L$-$Q$ and $R$-$Q$ tunnel couplings, the formal solution of Eq.~(\ref{aQ1}) can be written in the form
\begin{eqnarray}
a_Q(t) = a_Q^{(0)}(t) - \nonumber\\ \int dt_1 \langle -i [ a_Q^{(0)}(t), a_Q^{(0)\dag}(t_1)]_{+}\rangle \,\{ \Delta_L^* \,e^{-i\xi_{L}(t_1)}\,
a_L(t_1) + \Delta_R^*\, e^{-i\xi_{R}(t_1)} \,a_R(t_1)\}, \label{aQ2}
\end{eqnarray}
where $a_Q^{(0)}(t)$ is the free operator of the site $Q$, obeying the equation (\ref{aQ1}) with the tunneling terms neglected ($\Delta_L = 0,
\Delta_R = 0)$.

Taking into account the formula,
\begin{eqnarray}
i\Delta_L \,\langle e^{i\xi_{L}}\, a_L^\dag \,a_Q^{(0)} \rangle = \nonumber\\ |\Delta_L|^2\, \int dt_1 \langle
a_Q^{(0)\dag}(t_1)a_Q^{(0)}(t)\rangle \,\langle [e^{i\xi_{L}}(t)\, a_L^\dag (t),\, e^{-i\xi_{L}(t_1)} \,a_L(t_1)]_{+}\rangle\, \theta(t-t_1),
\label{aLaQ1}
\end{eqnarray}
which is similar to Eq.~(\ref{aLcS}), we obtain
\begin{eqnarray}
i\Delta_L \,\langle e^{i\xi_{L}}\, a_L^\dag \,a_Q \rangle = |\Delta_L|^2 \,\int dt_1 \{ \langle e^{-i\xi_{L}(t_1)} \,e^{i\xi_{L}(t)} \rangle
\langle a_Q^\dag (t_1) \,a_Q(t)\rangle \,\langle a_L(t_1) \,a_L^\dag (t)\rangle - \nonumber\\
\langle e^{i\xi_{L}(t)}\, e^{-i\xi_{L}(t_1)} \rangle\, \langle a_Q(t) \,a_Q^\dag (t_1)\rangle \,\langle a_L^\dag(t)\, a_L(t_1)\rangle \}.
\label{aLaQ2}
\end{eqnarray}
Dropping the label $^{(0)}$, we assume that the time evolution of the operators $a_Q$ in Eq.~(\ref{aLaQ2}) is calculated with the free-evolution
formula,
\begin{equation}
a_Q(t) = e^{-i\varepsilon_Q (t-t_1)}\, a_Q(t_1) - e^{- i\varepsilon_Q (t-t_1)}\, [ 1 - e^{i u_0 (t-t_1)} ] \,N_Q(t_1)\, a_Q(t_1). \label{aQ3}
\end{equation}
For the free-evolving proton operator of the interaction site we obtain a similar expression
\begin{equation}
b_Q(t) = e^{-iE_Q (t-t_1)}\, b_Q(t_1) - e^{- iE_Q (t-t_1)}\, [ 1 - e^{i u_0 (t-t_1)} ] \,n_Q(t_1)\, b_Q(t_1). \label{bQ1}
\end{equation}
The influence of the environment on the electron tunneling between the sites $L$ and $Q$, and between the sites $R$ and $Q$, is determined by
the correlators $\langle e^{-i\xi_{L}(t_1)}\, e^{i\xi_{L}(t)} \rangle$ and $\langle e^{i\xi_{L}(t)}\, e^{-i\xi_{L}(t_1)} \rangle$, where
\begin{equation}
\langle e^{i\xi_{L}(t)} \,e^{-i\xi_{L}(t_1)} \rangle = \exp\{-i\lambda_{L}(t-t_1)\} \,\exp\{-\lambda_{L}T(t-t_1)^2\}. \label{CorPhase}
\end{equation}
The reorganization energy, $\lambda_{L}$, is defined as \cite{Krish01}
\begin{equation}
\lambda_{L} = \sum_j \frac{m_j\omega_j^2}{2} (x_{jL} - x_{jQ})^2. \label{lambdaL}
\end{equation}
The electron reorganization energy $\lambda_{R}$, and the proton reorganization energies $\Lambda_{A}$ and $\Lambda_{B},$ are defined in a
similar way. In particular,
\begin{equation}
\Lambda_{A} = \sum_j \frac{m_j\omega_j^2}{2} (X_{jA} - X_{jQ})^2. \label{LambdaA}
\end{equation}

\subsubsection{Equations for populations of electron and proton-binding sites}
Consequently, we derive the system of rate equations for the average populations of the electron sites,
\begin{eqnarray}
\dot{n}_L + \gamma_S n_L = \gamma_S f_S (\varepsilon_L) + \Phi_L, \nonumber\\
\dot{n}_R + \gamma_D n_R = \gamma_D f_D (\varepsilon_R) + \Phi_R, \nonumber\\
\dot{n}_Q = - \Phi_L - \Phi_R, \label{nLRQ}
\end{eqnarray}
and for the average populations of the proton-binding sites,
\begin{eqnarray}
\dot{N}_A + \Gamma_N N_A = \Gamma_N F_N (E_A) + \Phi_A, \nonumber\\
\dot{N}_B + \Gamma_P N_B = \Gamma_P F_P (E_B) + \Phi_B, \nonumber\\
\dot{N}_Q = - \Phi_A - \Phi_B. \label{NABQ}
\end{eqnarray}
Here $\Phi_{\sigma}$ ($\sigma = L,R$) and $\Phi_{\alpha}$ ($\alpha = A,B$) are the functions of the average electron and proton populations,
respectively. In addition, due to a strong electron-proton Coulomb interaction on the site $Q$, the kinetic terms $\Phi_{\sigma}$ and
$\Phi_{\alpha}$ depend on the correlation function,
\begin{equation}
\langle K \rangle = \langle n_Q(t) N_Q(t)\rangle, \label{Kdef}
\end{equation}
 of the electron and proton populations on the site Q,
\begin{eqnarray}
\Phi_{\sigma} = \kappa_{\sigma }(\varepsilon_{\sigma} - \varepsilon_Q + \lambda_{\sigma }) \langle n_Q \rangle \langle 1 - n_{\sigma} \rangle -
\kappa_{\sigma }(\varepsilon_{\sigma} - \varepsilon_Q - \lambda_{\sigma }) \langle 1 - n_Q\rangle \langle n_{\sigma}
\rangle + \nonumber\\
\{ \kappa_{\sigma }(\varepsilon_{\sigma} - \varepsilon_Q + u_0 + \lambda_{\sigma }) - \kappa_{\sigma }(\varepsilon_{\sigma} - \varepsilon_Q +
\lambda_{\sigma })\} \langle 1 - n_{\sigma} \rangle \langle K
\rangle - \nonumber\\
\{ \kappa_{\sigma }(\varepsilon_{\sigma} - \varepsilon_Q + u_0 - \lambda_{\sigma }) - \kappa_{\sigma }(\varepsilon_{\sigma} - \varepsilon_Q -
\lambda_{\sigma })\} \langle n_{\sigma} \rangle \langle N_Q - K \rangle, \label{Phi1}
\end{eqnarray}
where $\kappa_{\sigma }(\varepsilon)$ is the Marcus rate for electron transfer between the site $\sigma$ and the interaction site $Q$,
\begin{equation}
\kappa_{\sigma }(\varepsilon) = |\Delta_{\sigma}|^2 \sqrt{\frac{\pi}{\lambda_{\sigma }T}}  \exp\left( - \frac{\varepsilon^2}{4 \lambda_{\sigma }
T } \right). \label{kappa1}
\end{equation}
The proton term $\Phi_{\alpha}$ is determined by the expression, similar to Eq.~(\ref{Phi1}), as
\begin{eqnarray}
\Phi_{\alpha} = \kappa_{\alpha }(E_{\alpha} - E_Q + \Lambda_{\alpha }) \langle N_Q \rangle \langle 1 - N_{\alpha} \rangle - \kappa_{\alpha
}(E_{\alpha} - E_Q - \Lambda_{\alpha }) \langle 1 - N_Q\rangle \langle N_{\alpha}
\rangle + \nonumber\\
\{ \kappa_{\alpha }(E_{\alpha} - E_Q + u_0 + \Lambda_{\alpha }) - \kappa_{\alpha }(E_{\alpha} - E_Q + \Lambda_{\alpha })\} \langle 1 -
N_{\alpha} \rangle \langle K
\rangle - \nonumber\\
\{ \kappa_{\alpha }(E_{\alpha} - E_Q + u_0 - \Lambda_{\alpha }) - \kappa_{\alpha }(E_{\alpha} - E_Q - \Lambda_{\alpha })\} \langle N_{\alpha}
\rangle \langle n_Q - K \rangle, \label{Phi2}
\end{eqnarray}
where $\kappa_{\alpha }(E)$ is the proton Marcus rate for the transitions between the site $\alpha$ and the proton-binding site $Q$,
\begin{equation}
\kappa_{\alpha }(E) = |\Delta_{\alpha}|^2 \sqrt{\frac{\pi}{\Lambda_{\alpha }T}}  \exp\left( - \frac{E^2}{4 \Lambda_{\alpha } T } \right).
\label{kappa2}
\end{equation}

\subsubsection{Equation for the electron-proton correlation function}
For the correlator, $\langle K \rangle$, of the electron ($n_Q$) and proton ($N_Q$) populations of the interaction site, we derive the following
equation
\begin{equation}
\langle \dot{K}\rangle = {\cal F}_L + {\cal F}_R +{\cal F}_A +{\cal F}_B, \label{K1}
\end{equation}
where
\begin{eqnarray}
{\cal F}_{\sigma} = \kappa_{\sigma }(\varepsilon_{\sigma} - \varepsilon_Q + u_0 - \lambda_{\sigma }) \langle n_{\sigma}\rangle
\langle N_Q - K \rangle - \nonumber\\
\kappa_{\sigma }(\varepsilon_{\sigma} - \varepsilon_Q + u_0 + \lambda_{\sigma }) \langle 1 - n_{\sigma}
\rangle \langle K\rangle, \nonumber\\
{\cal F}_{\alpha} = \kappa_{\alpha }(E_{\alpha} - E_Q + u_0 - \Lambda_{\alpha }) \langle N_{\alpha}\rangle \langle n_Q - K \rangle - \nonumber\\
\kappa_{\alpha }(E_{\alpha} - E_Q + u_0 + \Lambda_{\alpha }) \langle 1 - N_{\alpha} \rangle \langle K\rangle. \label{F1}
\end{eqnarray}

\subsubsection{Electron and proton currents}
Electron currents $I_S, I_D$ and proton currents $I_N,I_P$ are determined by an increase of the number of particles, electrons or protons, in
the corresponding reservoir. In particular, a variation of the electron number in the drain lead gives a current
\begin{equation}
I_D = \frac{d}{dt} \sum_k \langle c_{kD}^\dag c_{kD} \rangle = \gamma_D [ \langle n_R\rangle - f_D(\varepsilon_R) ], \label{ID}
\end{equation}
whereas the proton current $I_P$ is given by
\begin{equation}
I_P = \frac{d}{dt} \sum_q \langle d_{qP}^\dag d_{qP} \rangle = \Gamma_P [ \langle N_B\rangle - F_P(E_B) ]. \label{IP}
\end{equation}
Here,
$$\gamma_D = 2\pi \sum_k
|t_{kD}|^2 \delta(\varepsilon_R - \varepsilon_{kD}),$$ and
$$\Gamma_P = 2\pi
\sum_q |T_{qP}|^2 \delta(E_B - E_{qP})$$ are the electron ($\gamma_D$) and proton ($\Gamma_P$) transfer rates between the electron site $R$ and
the lead $D$, and between the proton-binding site $B$ and the $P$-side of the membrane, respectively. The multiplications of the particle
currents introduced above by the electron or proton charges produce the standard electric currents.

It follows from Eqs.~(\ref{nLRQ},\ref{NABQ}) that, in the steady-state, we have the relations: $$\langle \dot{n}_{\sigma} \rangle = 0\, ,\;
\langle \dot{N}_{\alpha} \rangle = 0\, ,$$ so that $$\Phi_L + \Phi_R = 0\,,$$ $$\Phi_A + \Phi_B = 0\, ,$$ and $$I_S = (d/dt)\sum_k \langle
c_{kS}^\dag c_{kS} \rangle = - I_D,$$ $$ I_N = (d/dt) \sum_q \langle d_{qN}^\dag d_{qN} \rangle = - I_P.$$

\subsubsection{Quantum yield of the electron-driven proton pump}
The productivity of the proton pump is determined by a quantum yield,
\begin{equation}
 QY = \frac{I_P}{I_D},
\label{QY}
\end{equation}
  and by the power-conversion efficiency $\eta$,
\begin{equation}
\eta = QY\times \frac{\mu_P-\mu_N}{\mu_S - \mu_D}. \label{eta}
\end{equation}
At the standard conditions, we have $$\mu_P - \mu_N = V_p + 60~{\rm meV} = 210~{\rm meV} ,$$ and $$\mu_S - \mu_D = V_e = 600~{\rm meV}  ,$$
therefore, $$\eta \simeq 0.35\times QY .$$ If a quantum yield $QY$ is of order one (or 100\%), the power-conversion efficiency $\eta$ may be as
much as 0.35 (or 35\%).

\subsection{Langevin equation}
For the redox-loop mechanism of a proton translocation through the membrane, the electron and proton sites, labelled by the letter $Q$, are
attached to the shuttle: a molecule diffusing between the $N$ and $P$ sides of the membrane (see Fig.~2). This Brownian motion can be described
by the one-dimensional overdamped Langevin equation for the coordinate $x$ of the shuttle,
\begin{equation} \zeta \dot{x} = -\, \frac{dU_{c}(x)}{dx} -
\Big\langle( n_Q - N_Q )^2\Big\rangle\,\frac{dU_{s}(x)}{dx} + \xi. \label{Langevin}
\end{equation}
We assume that the shuttle molecule moves along a line connecting the sites $L$ and $A$, located at $x=-x_0$, and the sites $R$ and $B$, both
having the coordinate $x=x_0$. The borders of the membrane, at $ x = \pm x_0,$ are schematically shown in Fig.~2. In Eq.~(\ref{Langevin}),
$\zeta$ is the drag coefficient of the shuttle, and $\xi$ is the Gaussian fluctuation force, which is characterized by the zero-mean value,
$\langle \xi\rangle = 0$, and the correlation function, $$\langle \xi(t)\xi(t')\rangle~=~2 \zeta T \delta (t-t')\, ,$$ proportional to the
temperature $T$ of the environment. The diffusion coefficient $D$ of the shuttle is also proportional to the temperature: $D = T/\zeta.$ The
motion of the shuttle is restricted by the membrane walls, which are simulated by the confinement potential $U_c(x)$,
\begin{equation}
U_c(x) = U_{c0} \left\{ 1  - \left[\exp\left( \frac{x - x_c}{l_c} \right) + 1 \right]^{-1} +   \left[ \exp\left( \frac{x + x_c}{l_c} \right) + 1
\right]^{-1} \right\}, \label{Uc}
\end{equation}
having the barrier height $U_{c0}$, the width $2x_c\ (x_c \geq x_0) $ and the steepness $l_c$.

The potential barrier $U_s(x)$,
\begin{equation}
U_s(x) = U_{s0} \left\{\left[\exp\left( \frac{x - x_s}{l_s} \right) + 1 \right]^{-1} -   \left[ \exp\left( \frac{x + x_s}{l_s} \right) +
1\right]^{-1} \right\}, \label{Us}
\end{equation}
does not allow the shuttle with a non-zero charge $q = N_Q - n_Q$ (in units of $|e|$)  to cross the lipid interior of the membrane. This barrier
is determined by the height $U_{s0}$, the steepness $l_s$, and the width $2x_s$.

\section{Results}
We solve the rate equations (\ref{nLRQ},\ref{NABQ}) for the electron ($n_{\sigma}$)  and proton ($N_{\alpha}$) populations jointly with the
equation (\ref{K1}) for the electron-proton correlation function on the site $Q$, $K = \langle n_Q N_Q \rangle$. Our approach can describe two
mechanisms of the redox-linked proton translocation across the membrane: (i) the static interaction site $Q$ and (ii) the situation when the
site $Q$ diffuses between the sides of the membrane. The mechanism (i) roughly corresponds to the proton pump operating in \emph{cytochrome c
oxidase} (CcO) \cite{WV07,BelPNAS07,Kim07,JCPCCO}, whereas the design (ii) can be attributed to the redox loop mechanism, which is responsible
for electron and proton transfers in the inner membrane of bacteria \cite{Mitch76,Jormakka02,Bertero03,Gennis05,PRERedoxLoop,JCPPhoto}.

\subsection{Static proton pump}
Here, we consider the mechanism (i), where the interaction site $Q$ does not change its position (see Fig.~1). We assume that protons are
transferred across the membrane, from the negatively charged side $N$, with an electrochemical potential $\mu_N$, to the positively charged side
$P$, having an electrochemical potential $\mu_P$. All potentials and energies are measured in meV.

\subsubsection{Parameters}
The difference of electrochemical potentials, $\Delta \mu_H = \mu_P - \mu_N,$ is determined by the following expression
\begin{equation}
\Delta \mu_H = V_p - 2.3 \,(RT/F)\times\Delta pH, \label{DMu}
\end{equation}
where $V_p$ is the transmembrane voltage, $R$ and $F$ are the gas and Faraday constants, respectively, $T$ is the temperature (in Kelvins, $k_B
= 1$), and the concentration gradient $\Delta pH$ is about $-1$.  \cite{Alberts02,Nicholls02}. The coefficient $2.3 \,(RT/F)$ is about 60~meV at
room temperature, $T = T_0 \equiv~298~$K. It follows from Eq.~(\ref{DMu}) that the potentials of the $N$ and $P$ sides of the membrane can be
written as
\begin{eqnarray}
\mu_N &=& -\mu_{H0} - \Delta V_p/2 - 30\times (\Delta T/T_0), \nonumber\\
\mu_P &=& \mu_{H0} + \Delta V_p/2 + 30\times (\Delta T/T_0), \label{muNP}
\end{eqnarray}
where $\Delta V_p = V_p - V_0,\; \Delta T = T - T_0. $ At the standard conditions, when $T = T_0$, $V_p = V_0 = 150$~meV, for the
electrochemical potential $\mu_{H0}$ we have: $\mu_{H0} = 105$~meV. Thus, the total proton gradient across the membrane, $\Delta \mu_H$, is
about 210~meV. As in the CcO proton pump \cite{WV07,JCPCCO}, we assume that the proton-binding sites $A,Q_p$, and $B$ are located approximately
on the line connecting the $N$ and $P$ sides of the membrane with the following coordinates: $x_A = 0.1,\;x_Q = 0.3,\; x_B = 0.5.$ The
coordinates of the sites are counted from the middle of the membrane in a direction towards the $P$-side and are measured in units of the
membrane width $W$ with $W\simeq 4$~nm. Protons are delivered from the $N$-side to the site $A$ by the so-called $D$-pathway crossing about a
half of the membrane. We also note that the $B$-site is located next to the $P$-side (see Fig.~1). An influence of the transmembrane voltage
$V_p$ on the energy levels of the proton sites is described by the formulas
\begin{eqnarray}
E_A = E_{A0} + x_A \times \Delta V, \nonumber\\
E_Q = E_{Q0} + x_Q \times \Delta V, \nonumber\\
E_B = E_{B0} + x_B \times \Delta V. \label{EAQB}
\end{eqnarray}
For the proton energy levels, $E_{A0}, E_{Q0}$, and $E_{B0}$, at the voltage $V_p = V_0$, we assume the following values (in meV): $E_{A0}=-155,
E_{Q0}=250$, and $E_{B0}=185,$ unless otherwise specified. This means that at the standard conditions, the proton begins its journey at the
$N$-side with the potential $\mu_N = - 105$~meV and jumps to the $A$-site having a lower energy ($-155$~meV). However, the next proton-binding
site $Q_p$ has a much higher energy ($\sim 250$~meV), so that the proton transfer cannot occur without a mediation of the electron component.
The electron site $Q_e$ is electrostatically coupled to the proton-binding site $Q_p$ with the Coulomb energy $u_0$. Thus, in the presence of an
electron on the site $Q_e$ the energy of the $Q$-proton decreases to the level $E_{Q0} - u_0 \simeq -220$~meV, provided that $u_0 \simeq
470$~meV. Now the proton can move from site $A$ to site $Q$, since $E_{A0} > E_{Q0} - u_0$. Depopulation of the electron site $Q$ returns the
energy level of the $Q$-proton to its original value $E_{Q0}=250$~meV, which is higher than the energy level of the next-in-line $B$ site,
$E_{B0}=185$~meV, and is \textit{much higher} than the energy level of the $A$-site. We assume that the backward proton transfer (from $Q_p$ to
$A$ site) is described by the \textit{inverted region} of the Marcus formula, so that the probability of such transfer is low, compared to the
probability of the proton transfer from the site $Q_p$ to the site $B$. No additional gate mechanism is necessary here.

For the sake of simplicity, we assume that three electron-binding sites $L,Q_e,R$ as well as the source and drain leads are positioned on a
line, which is parallel to the surface of the membrane (see Fig.~1). Thus, the transmembrane gradient $V_p$ has no effect on electron transport
from the electron source $S$ to the drain $D$. For the potentials of the electron reservoirs, we choose the following form
\begin{eqnarray}
\mu_S = \mu_{e0} + V_e/2, \nonumber\\
\mu_D = \mu_{e0} - V_e/2, \label{muSD}
\end{eqnarray}
with $\mu_{e0} = -500$~meV and with the electron voltage gradient $V_e = 600$~meV, unless otherwise indicated. The electron voltage gradient
$V_e$ roughly corresponds to the drop of the redox potential along the electron transfer chain in the cytochrome c oxidase
\cite{Alberts02,Nicholls02,WV07}. We assume that the electron pathway includes the source reservoir ($\mu_S = -200$~meV), the site $L$
($\varepsilon_L = -210$~meV), the interaction site $Q_e$ ($\varepsilon_Q = -250$~meV), the site $R$ ($\varepsilon_R = -770$~meV), and the
electron drain reservoir having the potential $\mu_D =-800$~meV.

We assume that the electron and proton transfer between the active sites, $L$-$Q$, $R$-$Q$ and $A$-$Q$, $B$-$Q$, are quite fast, with amplitudes
$\Delta_L \simeq \Delta_R \simeq 0.3$/ps and $\Delta_A \simeq \Delta_B \simeq 0.3$/ps, whereas the transitions to and out the electron and
proton reservoirs are characterized by much slower rates: $\gamma_S \simeq \gamma_D \simeq 1.5$/ns, and $\Gamma_N \simeq \Gamma_D \simeq
0.75$/ns. The responses of the environment to the electron and proton transitions are described by the corresponding reorganization energies:
$\lambda_L = \lambda_R = \lambda_e$ and $\Lambda_A = \Lambda_B = \Lambda_p,$ respectively. Here, for the standard case, we assume that
$\lambda_e \simeq 100$~meV and $\Lambda_p \simeq 100$~meV. This set of parameters provides an efficient operation of the redox-linked proton
pump.

\subsubsection{Dependence of the proton current on the transmembrane voltage}

In Fig.~3, we show the steady-state proton current $I_P$ as a function of the transmembrane voltage gradient $V_p$, at three different values of
the electron voltage: $V_e = 500,\; 600,\;700 $~meV. We use here the standard set of other parameters (see the previous subsection), where $T =
298$~K and $\lambda_e = \Lambda_p = 100$~meV.

The proton current $I_P$ is equal to the number of protons pumped \emph{energetically uphill} (at $V_p >0$), from the negative side $N$ to the
positive side $P$ of the membrane, \emph{per one microsecond}. At the difference $V_e = 600$~meV of source and drain redox potentials,  the
system pumps more than 200 protons per one microsecond against the transmembrane voltage gradient $V_p = 150$~meV. According to
Eq.~(\ref{muNP}), this voltage corresponds to the proton electrochemical gradient $\Delta \mu_H = 210$~meV, which is usually applied to the
internal membrane of mitochondria and the plasma membranes of bacteria. The number of pumped protons goes down as the proton voltage $V_p$
increases, and goes up with increasing the electron voltage difference $V_e$. The proton current saturates at $V_e > 750$~meV. It is evident
from Fig.~3 that at high enough electron voltages ($V_e \geq 600$~meV), the pump is able to translocate more than 100 protons per microsecond
against the proton gradient $V_p$, exceeding 250~meV ($\Delta \mu > 310$~meV). The quantum yield $QY$ is about one (with a power-conversion
efficiency $\eta \simeq 35$\%) in the whole region of electron and proton voltages:  500~meV $< V_e < 800$~meV, 0~$< V_p < 300$~meV.

\subsubsection{Proton current and the quantum yield as functions of temperature}
Figure~4 shows the pumping proton current, $I_P$ (i.e., the number of protons translocated from the negative to the positive side of the
membrane per one microsecond) versus the temperature $T$ measured in Kelvins. The graphs are presented at three values of the electron and
proton reorganization energy: $\lambda = 100,\;150,\;200$~meV. We assume here that $\lambda_e = \Lambda_p = \lambda$, with the electron voltage
$V_e = 600$~meV and the proton gradient $V_p = 150$~meV. It is of interest that at $\lambda \geq 150$~meV the pumping current has a pronounced
maximum near the room temperature, 200~K $ < T < 300$~K, although the quantum yield is higher, $QY \sim 1$, at lower temperatures. The
performance of the pump deteriorates at higher reorganization energies when the coupling to the environment increases. Increasing the
reorganization energy leads to increasing the probability for an electron to be transferred through the system, losing all its excess energy to
the environment without transferring this energy to protons. Such a probability is further increased at large temperatures leading to the
observed decrease of the quantum yield.

\subsubsection{Dependence of the proton current on the parameters of the interaction site}
The energy transfer from the electron to the proton component occurs on the interaction site $Q =\{Q_e,Q_p\}$, which has one electron
($\varepsilon_Q$) and one proton ($E_{Q}$) energy levels (see Eq.~(\ref{EAQB})). The electron on the site $Q_e$ is electrostatically coupled to
the proton, which populates the site $Q_p$, with the Coulomb energy $u_0$. It follows from Fig.~5 that the proton pumping current $I_P$ exhibits
a resonant behavior as a function of the  charging energy $u_0$ and the position of the proton energy level $E_{Q0}$. The dependence of the
pumping current on the electron energy $\varepsilon_Q$ has a resonant character as well. Here we assume that $V_e = 600$~meV, $V_p = 150$~meV,
$\lambda_e = \Lambda_p = 100$~meV, and $T = 298$~K. The energetically-uphill proton current has a pronounced maximum ($I_P \simeq 220/\mu$s) at
the Coulomb energy $u_0 = 470$~meV and the proton energy $E_{Q0} = 250$~meV, provided that the electron energy $\varepsilon_Q = -250$~meV. It is
important that the proton pump is robust to the variations of the Coulomb energy $u_0$ and the proton energy $E_{Q0}$ in the range $\pm 50$~meV
from the resonant values. The quantum yield $QY$ is very close to one in the central region of Fig.~5, so that the power-conversion efficiency
$\eta$ is about of 35\%.

Figures 3, 4 and 5 clearly demonstrate that, at standard physiological conditions, the static redox-linked proton pump (``CcO-pump") efficiently
converts the energy of electrons to the more stable energetic form of the proton electrochemical gradient across the membrane.

\subsection{Redox loop mechanism of electron and proton translocation}

In many biological systems, electrons and protons can be transferred across a membrane by means of a molecular shuttle diffusing inside of the
membrane, from one side to another. Here we show that the mathematical model described in Section II can be successfully applied for a
description of the redox loop mechanism, which utilizes the Brownian motion of the shuttle $Q$ carrying both electron, $Q_e$, and proton, $Q_p$,
sites (see Fig.~2). As in the previous case, we have to solve here a system of master equations for the electron ($n_L, n_Q, n_R$) and proton
($N_A, N_Q, N_B$) populations, Eqs.~(\ref{nLRQ},\ref{NABQ}), and for the correlation function $K$ of electron and proton populations on the site
$Q$, Eq.~(\ref{K1}). However, these master equations should be complemented by the Langevin equation, Eq.~(\ref{Langevin}), for the
time-dependent shuttle position $x$. We note that the electron tunneling between the sites $L$-$Q$, $Q$-$R$, as well as the proton transfer
rates between the sites $A$-$Q$ and $Q$-$B$, depend on the position $x$ of the shuttle.

\subsubsection{Parameters}
We assume that the electron site $L$ is located near the negative ($N$) side of the membrane, at $x = - x_0$, where $x_0 = 2$~nm. The other
electron site $R$ is near the $P$-side of the membrane, at $x = + x_0$. The reservoir $S$, connected to the site $L$, serves as a source of
electrons, and the reservoir $D$, coupled to the site $R$, serves as an electron drain (see Fig.2). The tunneling amplitudes $\Delta_L,
\Delta_R$ are determined by the amplitudes $\Delta_{L0},\;\Delta_{R0}$, and by the electron tunneling length $l_e$:
\begin{eqnarray}
\Delta_L(x) = \Delta_{L0}\times \exp\left(- \frac{|x+x_0|}{l_e}\right), \nonumber\\
\Delta_R(x) = \Delta_{R0}\times \exp\left(- \frac{|x-x_0|}{l_e}\right). \label{DeltaLR}
\end{eqnarray}
The proton-binding site $A$ is located at the end of the $N$-side proton pathway, whereas the site $B$ terminates a pathway, which goes into the
$P$-side of the membrane. For the $x$-dependencies of the proton transfer amplitudes $\Delta_A$ and $\Delta_B$, we choose the following
relations:
\begin{eqnarray}
\Delta_A(x) = \Delta_{A0}\times \left[\exp\left(\frac{x_0+x}{l_p}\right) + 1 \right]^{-2}, \nonumber\\
\Delta_B(x) = \Delta_{B0}\times \left [\exp\left(\frac{x_0-x}{l_p}\right) + 1 \right]^{-2}, \label{DeltaAB}
\end{eqnarray}
where $l_p$ is the proton transfer length. It should be noted that our model produces the same results when the proton amplitudes are given by
the expressions similar to Eqs.~(\ref{DeltaLR}). For the transfer parameters, we choose the following values: $\Delta_{L0} \sim \Delta_{R0} =
0.04$~meV, $\Delta_{A0} \sim \Delta_{B0} = 0.04$~meV, and $l_e = 0.25$~nm, $l_p = 0.25$~nm. Couplings to the electron and proton reservoirs are
described by the rates $\gamma_S \sim \gamma_D = 0.5/$ns and $\Gamma_N \sim \Gamma_P = 0.1/$ns. The system is robust to significant variations
of the transfer parameters.

The confinement potential $U_c(x)$ is determined by the height $U_c = 500$~meV, the steepness $l_c = 0.1$~nm, and the half-width $x_c = 2.7$~nm.
The potential barrier $U_s(x)$, preventing the charged shuttle from entering into the membrane, is characterized by the height $U_s = 770$~meV,
the width $x_s = 1.7$~nm, and the steepness $l_s = 0.05$~nm.

Accordingly, the electron and proton populations of the shuttle are almost completely compensated, $n_Q \simeq N_Q$, so that the potential
$U_s(x)$ gives a negligible contribution to the energies of electrons and protons. However, we have to take into account the fact that in the
presence of the voltage gradient, $V_p \simeq 150$~meV, the electron ($\varepsilon_Q$) and proton ($E_Q$) energies on the moving shuttle depend
on the shuttle  position $x$:
\begin{eqnarray}
\varepsilon_Q = \varepsilon_{Q0} - \frac{x}{2 x_0} V_p, \nonumber\\
E_Q = E_{Q0} + \frac{x}{2 x_0} V_p,
\end{eqnarray}
with $\varepsilon_{Q0} = 280$~meV, and $E_{Q0} = u_0/2 = 200$~meV, where for the charging energy $u_0$ of the shuttle we have: $u_0 = 400$~meV.

Thus, electrons move from the source reservoir, having the electrochemical potential $\mu_S = 420$~meV, to the $L-$site (with the energy
$\varepsilon_L = 380$~meV), and, thereafter, to the shuttle. On the opposite side of the membrane, the electron, populating the shuttle, jumps
to the site $R$ ($\varepsilon_R = -170$~meV) and, finally, to the drain reservoir ($\mu_D = - 230$~eV). The total drop of the redox potential in
this electron-transport chain can be estimated as $\mu_S - \mu_D = 650$~meV.

Protons move from the $N$-side of the membrane ($\mu_N = - 105$~meV) to the site $A$, having a lower energy $E_A = -150$~meV. The energy level
$E_Q = 125$~meV of the proton on the shuttle, located near the $N$-side of the membrane (at $x = - x_0$), is much higher than $E_A$, if the
shuttle contains no electrons. However, the shuttle populated with a single electron is more attractive for protons, since in this case the
effective energy of the proton, $E_Q - u_0 = -275$~meV, is less than the energy of the proton-binding site $A$. The shuttle, carrying one
electron and one proton, diffuses to the opposite side of the membrane ($x = + x_0$), where the electron, with energy $\varepsilon_Q - u_0 = -
195$~meV, is able to tunnel to the site $R$, having a slightly higher energy $\varepsilon_R = - 170$~meV. In the absence of an electron, the
energy of the proton on the shuttle (at $x = + x_0$) increases to the level $E_Q = E_{Q0} + V_p/2 = 275$~meV, which exceeds the energy of the
proton on the site $B$: $E_B = 150$~meV. Consequently, the proton moves from the shuttle to the site $B$ and, thereafter, to the $P$-side of the
membrane characterized by the electrochemical potential $\mu_P = + 105$~meV. Thus, this redox loop mechanism translocates protons across the
membrane against the proton electrochemical gradient $\Delta \mu_H = \mu_P - \mu_N = 210$~meV, and against the transmembrane potential $V_p \sim
150$~meV.

\subsubsection{Proton translocation process}
Figure~6 exhibits the electron and proton populations of the shuttle, $n_Q(t)$ and $N_Q(t)$, correlated with the shuttle's position $x(t)$ at $T
= 298$~K, $V_p = 150$~meV, and at $\Delta \mu = 210$~meV. In this figure, we also show the time dependencies of the number of electrons,
$n_D(t)$, transferred to the drain reservoir, and the number of protons, $N_P(t)$, translocated to the positive side of the membrane. The
shuttle diffuses between the membrane walls located at $x = \pm x_0$  ($x_0 = 2$~nm) with an average crossing time $\Delta t \sim 2.5\,\mu$s.
This time-scale is closely related to the diffusion time, $$t_D \sim \langle \Delta x^2 \rangle /2D \sim 2.66 \;\mu{\rm s}\, ,$$ obtained at
$\sqrt{\langle \Delta x^2 \rangle} \sim 2 x_0 = 4$~nm, for the diffusion coefficient of the quinone molecule $D \sim 3\cdot
10^{-12}\,$m$^2/$sec.

At $t \sim 0$, the shuttle, located at $x \sim - x_0$, is loaded with one electron and one proton taken from the negative side of the membrane
(see Fig.~2). When $t \sim 2.5\;\mu$s, the shuttle reaches the positive side ($x = + x_0 = 2$~nm) and unloads the electron to the the site $R$
(and later to the drain lead $D$) and the proton to the site $B$, coupled to the $P$-side of the membrane. Consequently, the population $N_P$ of
the $P$-side grows. The empty shuttle diffuses back, to the $N$-side, completing the cycle, and the process starts again. In twenty
microseconds, the shuttle performs four complete trips and translocates about four electrons and four protons across the membrane.

\subsubsection{Voltage and temperature dependencies}
The numbers of electrons and protons, $n_D$ and $N_P$, respectively, transferred across the membrane in one millisecond, are shown in Fig.~7 as
functions of the transmembrane proton voltage $V_p.$ The electrochemical gradient of protons, $\Delta \mu = \mu_P - \mu_N$, is proportional to
$V_p$: $\Delta \mu~\simeq~V_p~+~60$~meV (at $T = 298$~K). The results in Fig.~7 are averaged over ten realizations. The system is able to
translocate more than 120 protons per ms against the high transmembrane voltage, $V_p \leq 250$~meV, that corresponds to the electrochemical
gradient $\Delta \mu \leq 310$~meV.

It follows from Fig.~8 that the translocation mechanism works efficiently in a wide range of temperatures, 250~K $< T < 500$~K. In this range,
the system pumps more than 120 protons per millisecond with a quantum yield exceeding 90\% and with a power-conversion efficiency $\eta$ higher
than 40\%. With increasing temperature, the shuttle performs more trips between the sides of the membrane, thus, carrying more electrons and
protons. This increases the proton current (i.e., the number of protons translocated per unit time). We note that the proton population of the
shuttle occurs only after loading the shuttle with an electron. At very high temperatures, $T > 500$~K, the shuttle moves quite fast, and
protons have less chances to jump on the shuttle. Consequently, the gap between electron and proton currents grows with the temperature, thus
deteriorating the performance of the pump.

\section{Conclusion}

Two different mechanisms of energetically-uphill proton translocation across a biomembrane are described by the same physical model. This model
includes three redox sites $(L, Q_e, R)$ and three proton binding sites $(A, Q_p, B)$ attached to the source ($S$) and drain ($D$) electron
reservoirs, as well as to the proton reservoirs on the positive and negative sides of the membrane. We have shown that it is the strong Coulomb
interaction between the electron site $Q_e$ and the proton site $Q_p$, which plays the most prominent role in the process of energy
transformation from electrons to protons. In this case, the whole electron-proton transport chain can be divided into weakly coupled clusters of
sites, so that the total number of occupation states is equal to the sum (not to the product) of occupation states in each cluster. At
physiological conditions, our model demonstrates a proton pumping effect with a quantum yield near 100\% and a power-conversion efficiency of
order of 35\%, for both the static proton pump, related to the \emph{cytochrome c oxidase}, as well as for the redox-loop mechanism, where
electrons and protons are translocated by the diffusing molecular shuttle.

\textbf{Acknowledgements.} This work was supported in part by the Laboratory of Physical Sciences, National Security Agency, Army Research
Office, National Science Foundation grant No. 0726909, JSPS-RFBR contract No. 09-02-92114, Grant-in-Aid for Scientific Research (S), MEXT
Kakenhi on Quantum Cybernetics, and Funding Program for Innovative R\&D on S\&T (FIRST). L.M. was partially supported by the NSF NIRT, Grant No.
ECS-0609146 and by the PSC-CUNY Award No. 41-613.

\begin{figure}
\includegraphics[width=14.0cm, height=16.0cm ]{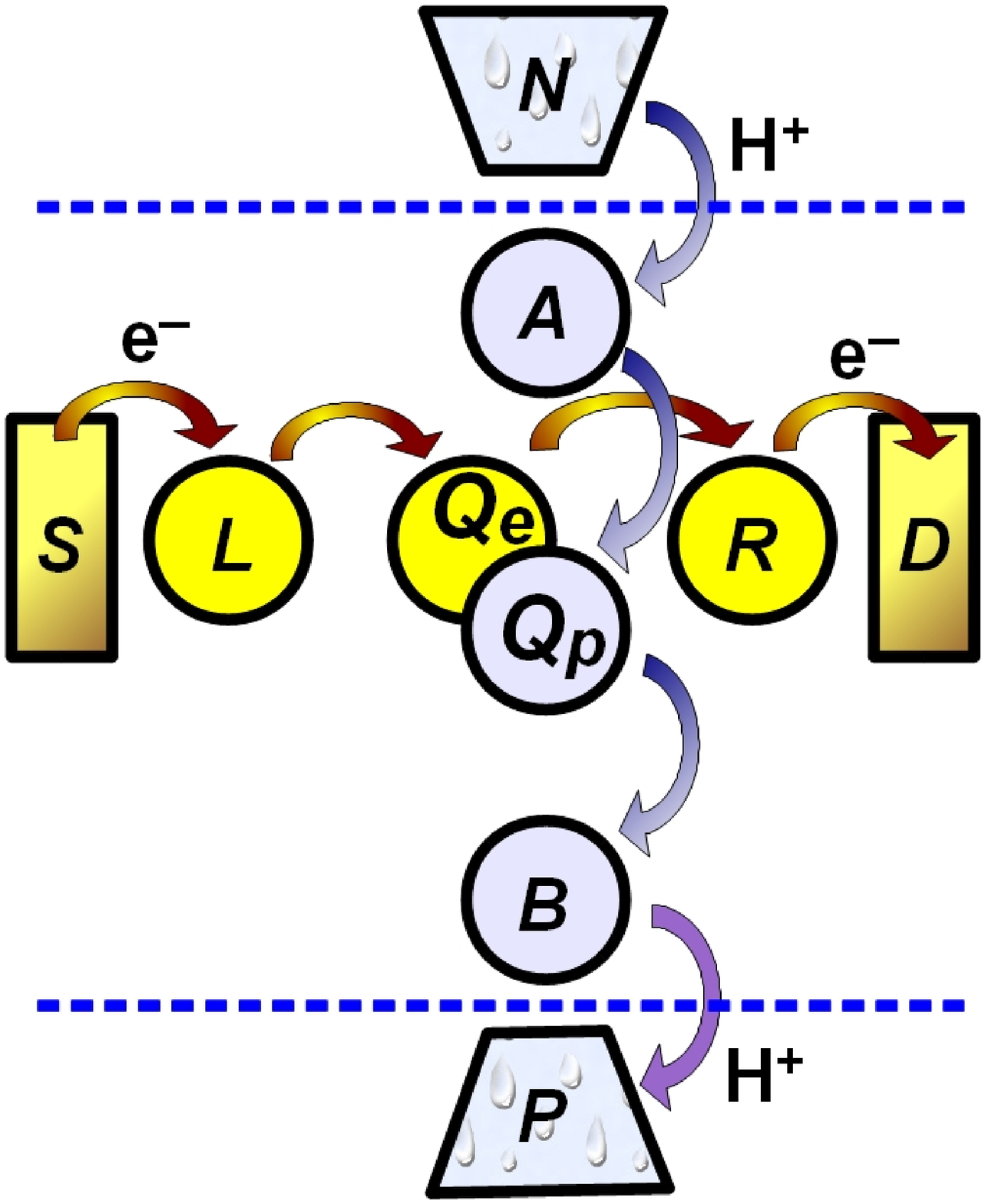}
\vspace*{0cm} \caption{ (Color online) Schematic diagram of the static proton pump. The electron transport chain starts at the source ($S$)
lead. Thereafter, high-energy electrons, ${\rm e}^-$, tunnel \emph{energetically-downhill} (through the yellow path) to the sites $L,\;Q_e,\;R$
and, finally, to the drain $D$. Low-energy protons, ${\rm H}^+$, move \emph{energetically-uphill} (in blue) from the negative ($N$) side of the
membrane to the sites $A, \;Q_p,\;B$ and, eventually, reach the positive ($P$) side of the membrane.  }
\end{figure}

\begin{figure}
\includegraphics[width=11.0cm, height=13.0cm ]{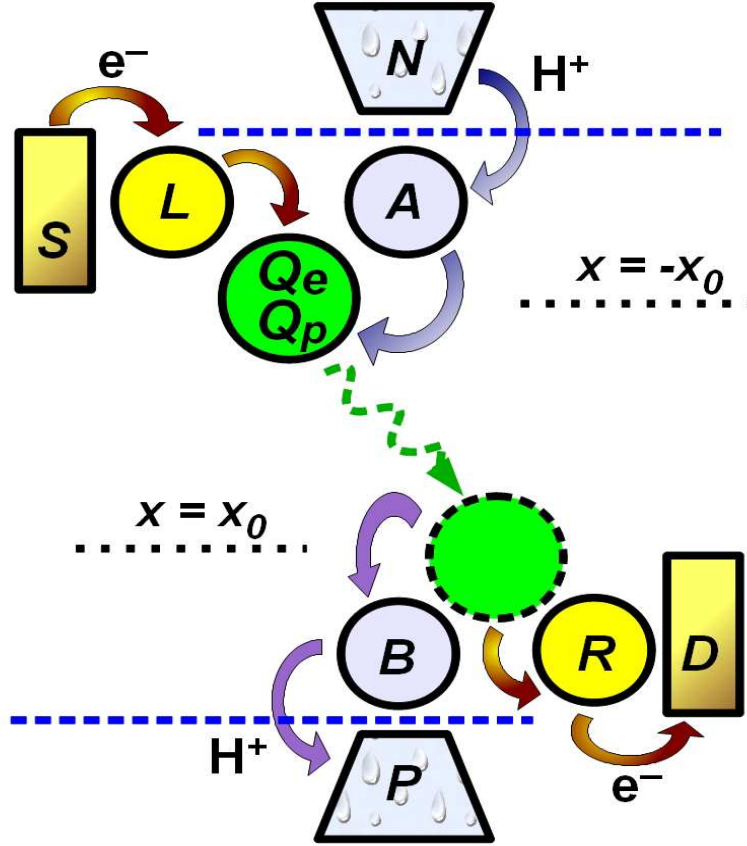}
\vspace*{0cm} \caption{ (Color online) Schematic diagram of the redox loop mechanism. Here, the electron-proton interaction site, $Q =
\{Q_e,Q_p\}$, is placed on the molecular shuttle (shown in green), which diffuses along the line connecting the negative and positive sides of
the membrane. From the source reservoir $S$, an electron ${\rm e}^-$ jumps to the site $L$ and, thereafter, to the shuttle, located at $x = -
x_0.$ The shuttle also accepts a proton ${\rm H}^+$ transferred from the $N$-side of the membrane via the site $A$. The loaded shuttle moves
randomly toward the positive side ($P$) of the membrane, where (at $x = x_0$) the electron is subsequently transferred from the site $Q_e$ to
the site $R$ and to the drain reservoir $D$, and the proton jumps from the site $Q_p$ to the site $B$ and, finally, to the positive ($P$) side
of the membrane. We note that, in this design, the electron site $L$ and the proton site $A$ are located near the $N$-side of the membrane
(shown by the horizontal blue dashed line), and the electron site $R$ and the proton site $B$ are placed near the $P$-side. }
\end{figure}

\begin{figure}
\includegraphics[width=21.0cm, height=12.0cm ]{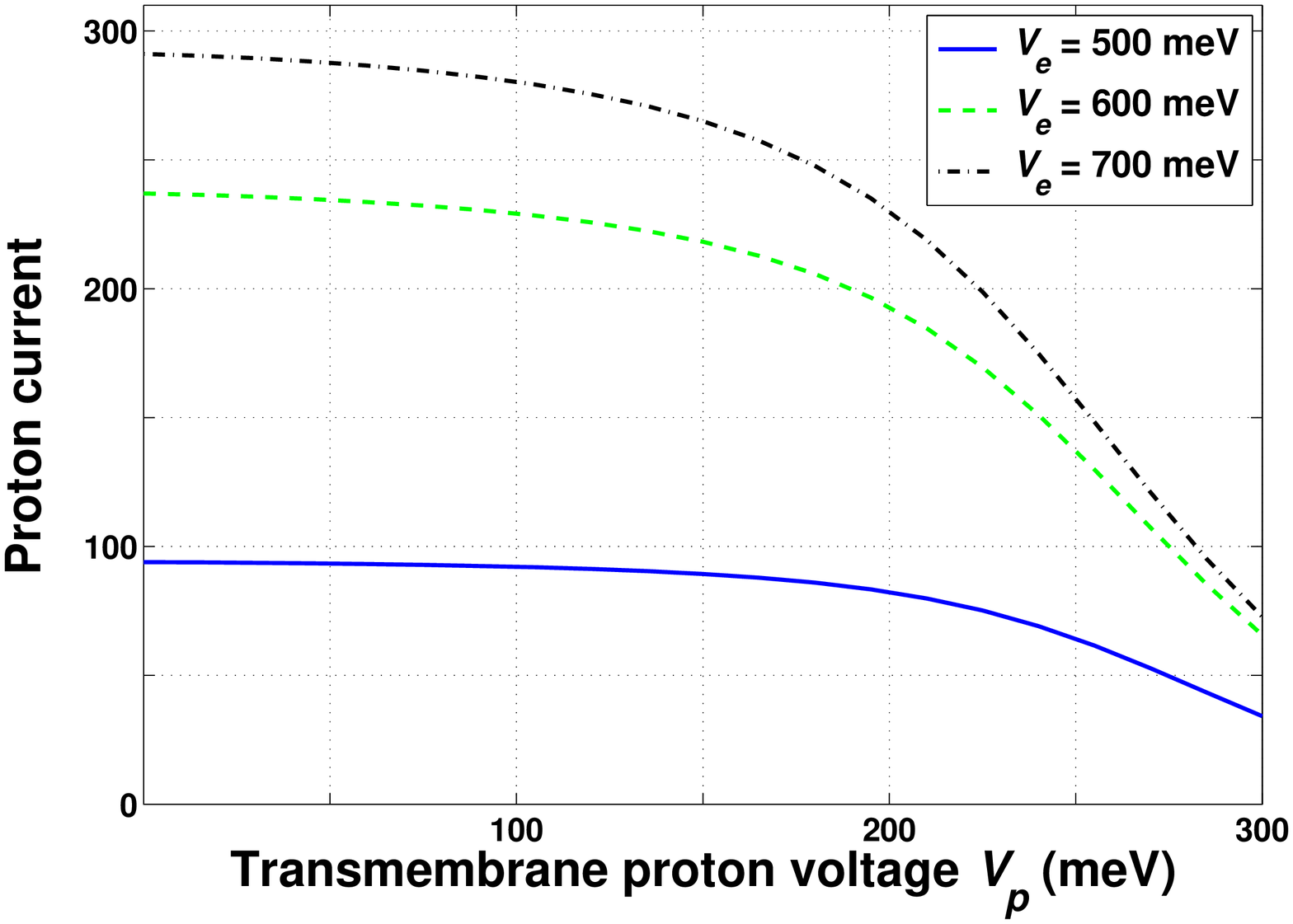}
\vspace*{0cm} \caption{ (Color online) Proton current versus transmembrane voltage $V_p$ at room temperature, $T = 298$~K,  and three different
electron potentials: $V_e = 500,\; 600,\;$ and $700$~meV. The proton current is almost constant for low values of $V_p$, and decreases for
increasing $V_p$. }
\end{figure}

\begin{figure}
\includegraphics[width=21.0cm, height=12.0cm ]{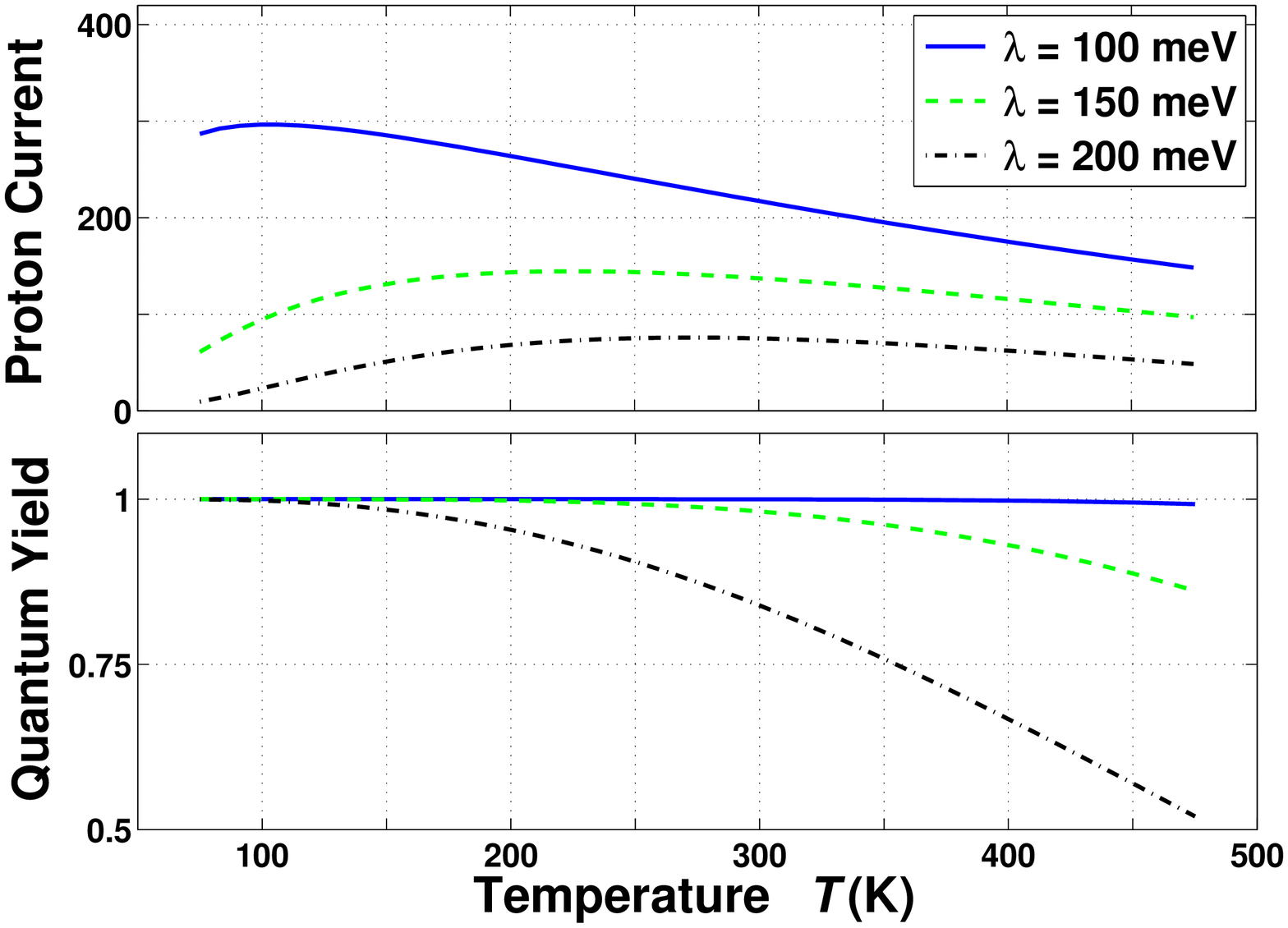}
\vspace*{0cm} \caption{ (Color online) Proton current (the number of protons translocated across the membrane per one microsecond) and quantum
yield versus temperature for the electron voltage $V_e = 600$~meV, transmembrane proton voltage $V_p = 150$~meV, and three different
reorganization energies: $\lambda = 100,\;150,\;$ and $200$~meV.  The proton current and quantum yield both decrease, for increasing $\lambda
$.}
\end{figure}

\begin{figure}
\includegraphics[width=18.5cm, height=12.0cm ]{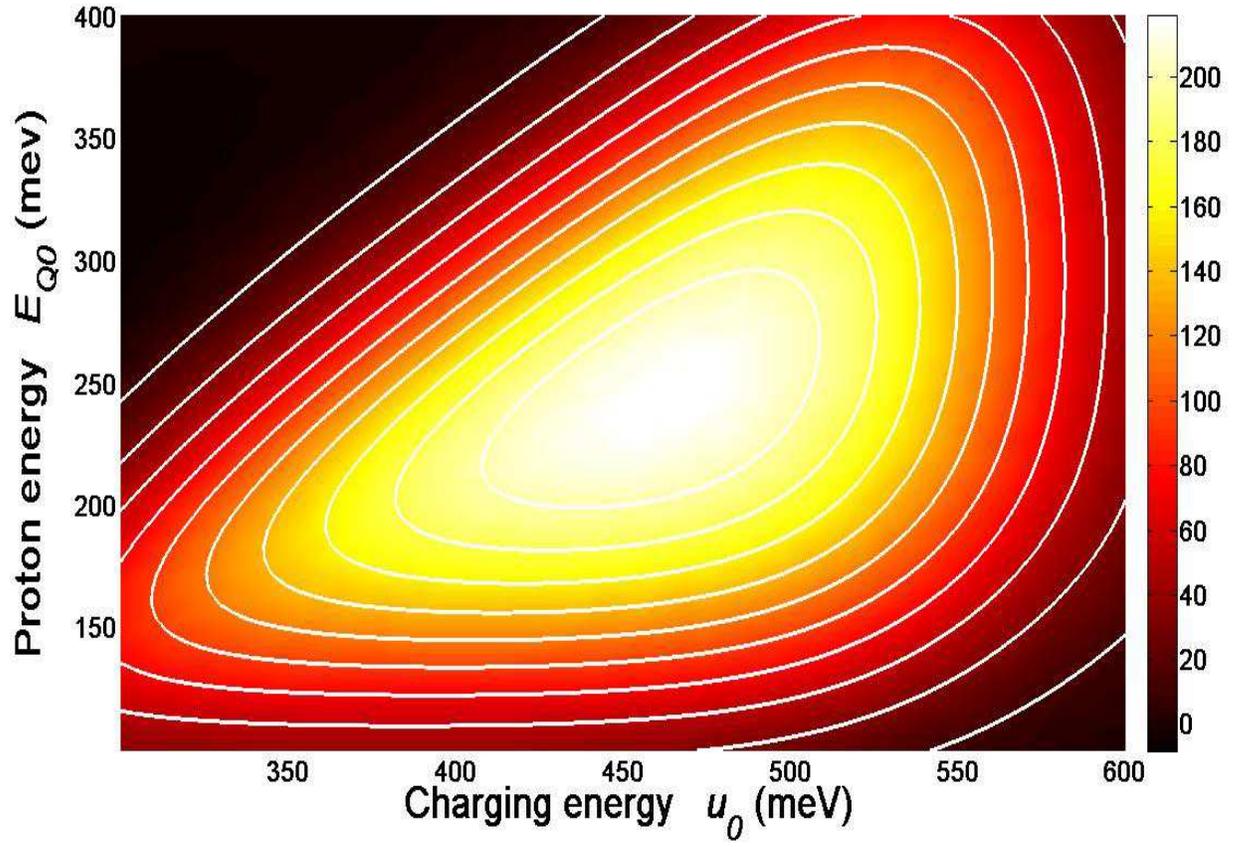}
\vspace*{0cm} \caption{ (Color online) Dependence of the proton current (the number of protons pumped across the membrane per  one $\mu$s, see
the color bar on the right side) on the charging energy $u_0$, and on the energy $E_{Q0}$ of the central proton site for  $V_e = 600$~meV, $V_p
= 150$~meV, and $T = 298$~K. }
\end{figure}

\begin{figure}
\includegraphics[width=22.0cm, height=12.0cm ]{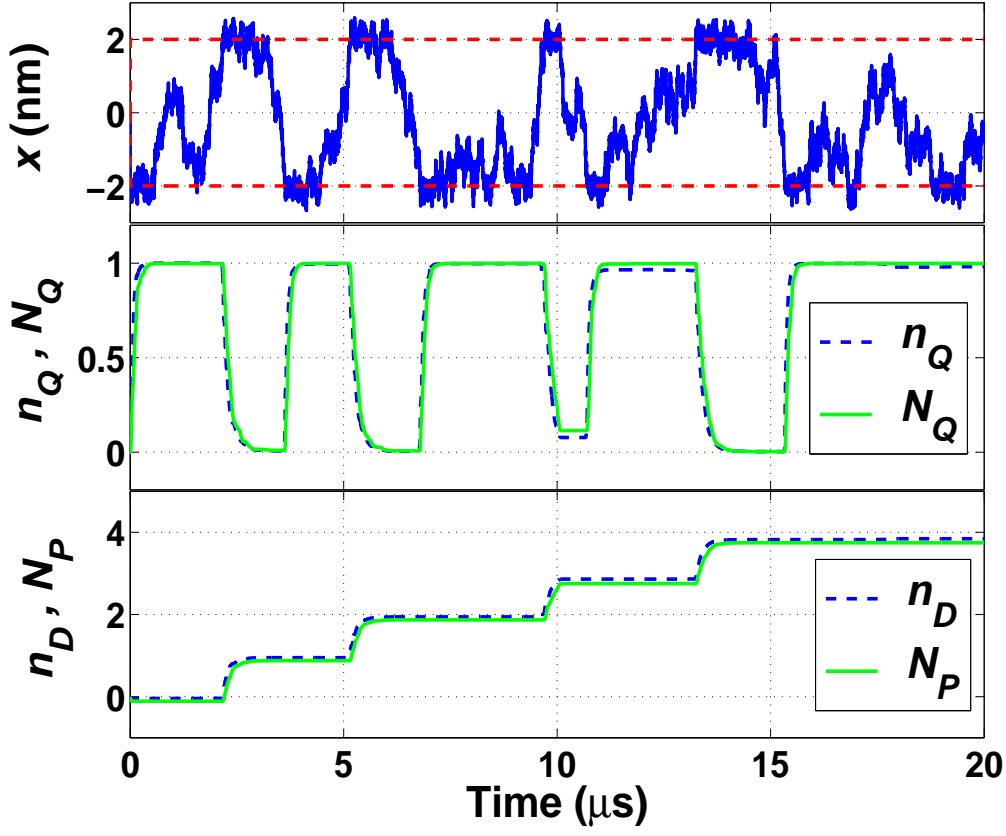}
\vspace*{0cm} \caption{ (Color online) Time evolution of the electron-proton translocation process. Here $x$ is the location of the shuttle,
$n_Q$ and $N_Q$ are the electron and proton populations of the shuttle, respectively, $n_D$ is the number of electrons transferred from the
electron source $S$ to the electron drain $D$, and $N_P$ is the number of protons translocated from the negative ($N$) to the positive ($P$)
side of the membrane. It can be seen from this figure that the loading/unloading of the shuttle with electrons and protons, as well as the
electron and proton transfer across the membrane, are clearly correlated with the spatial motion of the shuttle. }
\end{figure}

\begin{figure}
\includegraphics[width=18.0cm, height=10.0cm ]{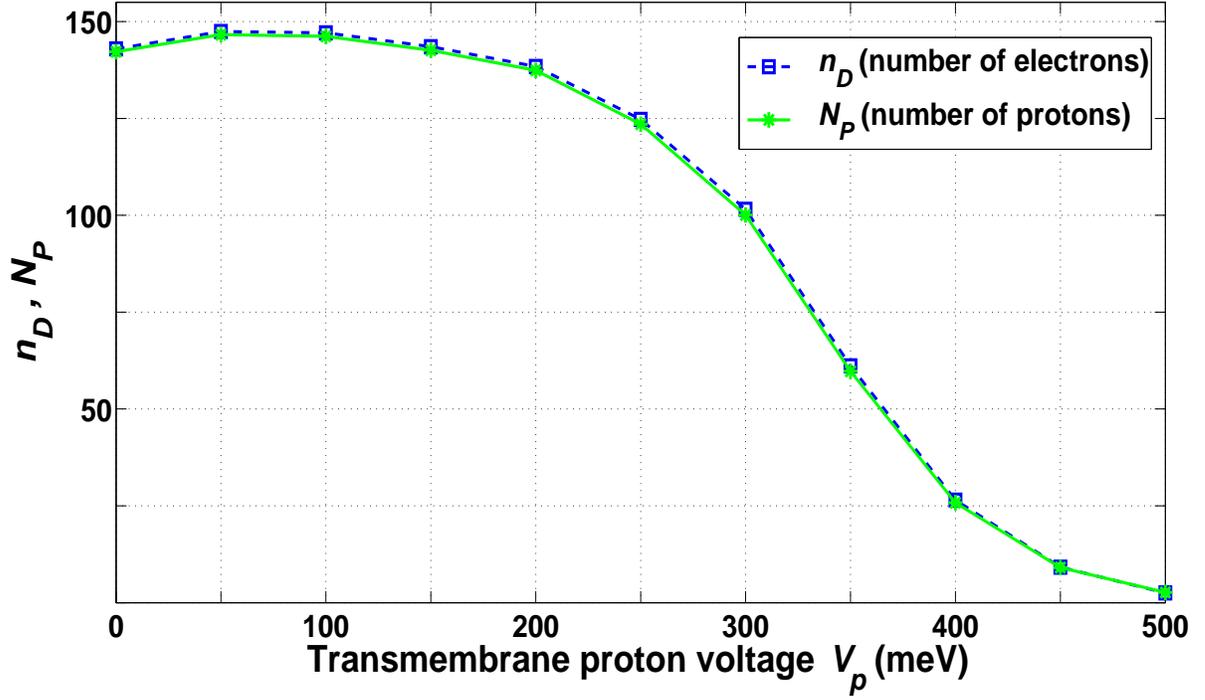}
\vspace*{0cm} \caption{ (Color online) Numbers of electrons, $n_D$, and protons, $N_P$, translocated across the membrane in one millisecond,
versus the transmembrane proton voltage $V_p$ at room temperature, $T = 298$~K, and at $(\mu_S - \mu_D) = 650$~meV. Clearly, it is much harder
to transfer protons against the higher transmembrane voltages. }
\end{figure}

\begin{figure}
\includegraphics[width=22.0cm, height=12.0cm ]{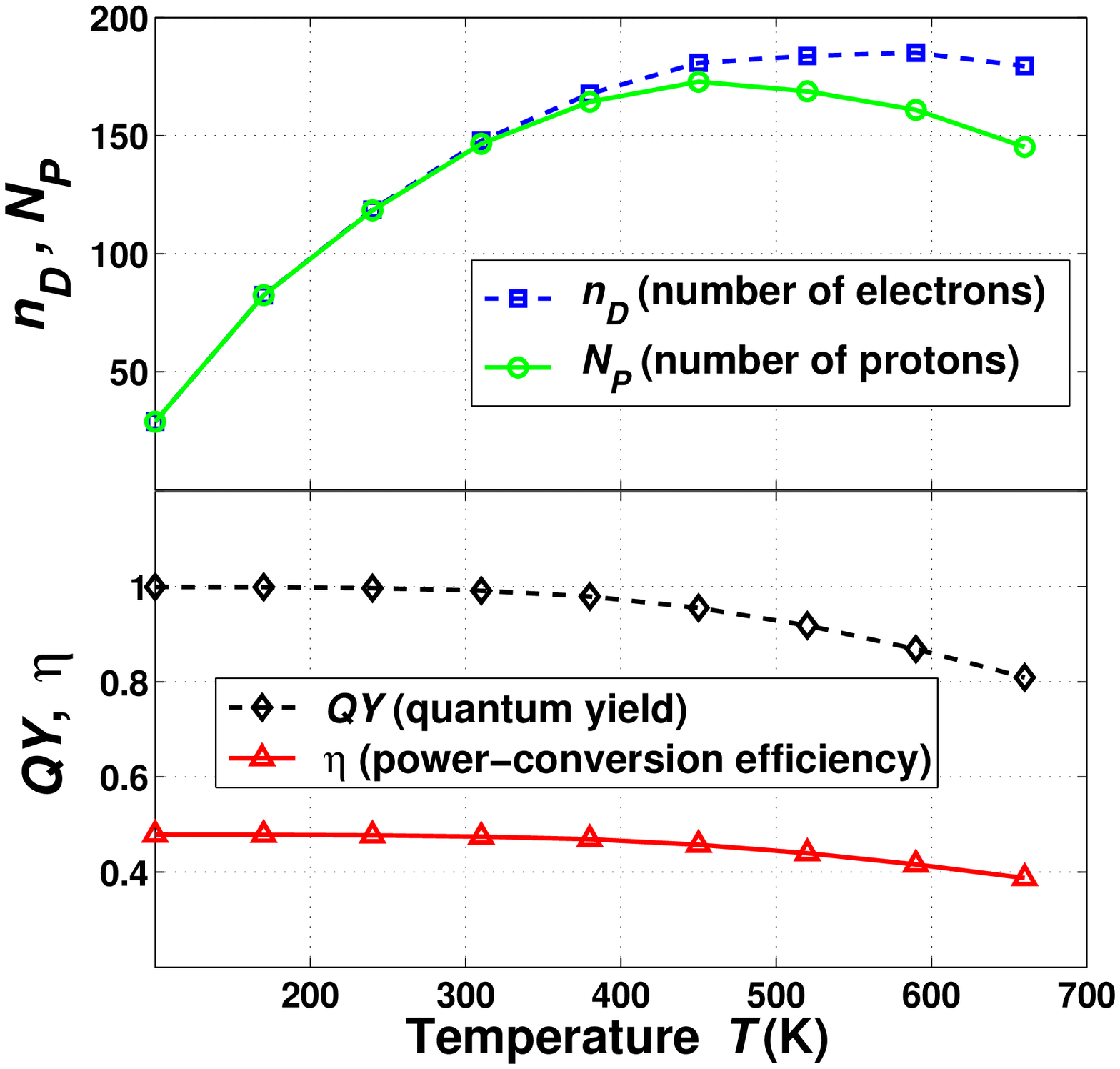}
\vspace*{0cm} \caption{ (Color online) Temperature dependence of the numbers of electrons, $n_D$, and protons, $N_P$, transferred across the
membrane by the diffusing shuttle, at $V_p = 150$~meV and $(\mu_S - \mu_D) = 650$~meV. We also present here the quantum yield, $QY$, and the
power-conversion efficiency, $\eta$, of the process as functions of the temperature. At higher temperatures, the shuttle moves faster and
carries more electrons and more protons. However, if the temperature is too high, the shuttle has not enough time to be loaded with electrons
and protons, and sometimes travels empty. As a result of this, the electron and proton currents decrease at high temperatures, thus decreasing
the efficiency of the pump. }
\end{figure}

\end{document}